\documentclass[aps,prd,twocolumn,superscriptaddress,floatfix,nofootinbib]{revtex4-2}

\usepackage[utf8]{inputenc}
\usepackage[T1]{fontenc}
\usepackage[english]{babel}
\usepackage{amsmath,amssymb,bm}
\usepackage{mathtools}
\allowdisplaybreaks[3]
\usepackage{microtype}
\usepackage{graphicx}
\usepackage{booktabs}
\usepackage{xcolor}
\usepackage[colorlinks=true,linkcolor=blue,citecolor=blue,urlcolor=blue]{hyperref}

\newcommand{\rs}{r_s}

\newcommand{\epse}{\epsilon_{\rm e}}
\newcommand{\Mpl}{M_{\rm Pl}}
\newcommand{\Oe}{\mathcal O_{\rm e}}
\newcommand{\Oo}{\mathcal O_{\rm o}}

\newcommand{\dd}{\mathrm d}
\newcommand{\cO}{\mathcal O}
\newcommand{\Rfix}{\mathcal R_{2}^{\rm fix}}
\newcommand{\kfix}{k_{2,{\rm sc}}^{\rm fix}}
\newcommand{\aone}{\alpha_1}

\begin{document}

\title{Fixed-quadrupole static tidal response of Schwarzschild black holes in a cubic Weyl effective field theory}

\author{Edilberto O. Silva}
\email{edilberto.silva@ufma.br}
\affiliation{Programa de P\'{o}s-Gradua\c{c}\~{a}o em F\'{i}sica, Universidade Federal do Maranh\~{a}o, 65080-805, S\~{a}o Lu\'{i}s, Maranh\~{a}o, Brazil}
\affiliation{Coordena\c{c}\~ao do Curso de F\'{i}sica -- Bacharelado, Universidade Federal do Maranh\~{a}o, 65085-580, S\~{a}o Lu\'{i}s, Maranh\~{a}o, Brazil}

\date{\today}

\begin{abstract}
Static Love numbers of four-dimensional Schwarzschild black holes vanish in general relativity. We study how the fixed-quadrupole static tidal solution is modified by the parity-even cubic Weyl operator in the gravitational effective field theory. Working perturbatively in $\epse=\lambda_{\rm e}(\Lambda\rs)^{-4}$, we construct the reduced quadratic radial action for static even-parity $\ell=2$ perturbations, order-reduce the higher-derivative equations, and solve the resulting boundary-value problem directly in metric variables. The order-$\epse$ equations reduce to a first-order two-dimensional inhomogeneous system for $X_0$ and $X_K$, with $X_2$ fixed by an algebraic constraint. Horizon regularity leaves one constant, but matching to infinity shows that this freedom only renormalizes the applied tidal branch. After removing this tidal renormalization, the decaying branch is unambiguous. Calibrating the spatial sector at fixed $\ell=2$ against the associated-Legendre branches $P_2^{\,2}$ and $Q_2^{\,2}$, we obtain
$\Delta(B/A)_{\ell=2}^{\rm fixed}=-2400\,\epse,
\; \Delta k_{2,{\rm sc}}^{\rm fix}=-20\,\epse .$
The second number is a scalar fixed-$\ell$ conversion of the metric branch ratio, not yet analytically continued, gauge-invariant electric Love number. A comparison with canonical Teukolsky-based Love numbers requires an additional continuation in $\ell$ and a precise map of normalizations. The result should therefore be viewed as a reproducible metric-sector benchmark for the cubic Weyl EFT, complementary to gauge-invariant master-equation approaches.
\end{abstract}

\maketitle

\section{Introduction}
\label{sec:introduction}

Tidal Love numbers quantify the conservative response of a compact object to an external gravitational field~\cite{Love1909}.  They are central to the effective description of compact binaries because they multiply local worldline operators and determine finite-size corrections to conservative dynamics and to gravitational waveforms~\cite{FlanaganHinderer2008,Hinderer2008,YagiYunes2013,CardosoFranzinMaselliPaniRaposo2017}.  For ordinary stars, these coefficients encode the equation of state and internal structure.  Four-dimensional black holes in general relativity (GR), by contrast, have vanishing static Love numbers~\cite{BinningtonPoisson2009,DamourNagar2009,HuiJoycePencoSantoniSolomon2021,Chia2021,LeTiecCasalsFranzin2021}.  This cancellation is now understood from several viewpoints, including direct black-hole perturbation theory, effective field theory (EFT), and hidden symmetry structures of static perturbations~\cite{GoldbergerRothstein2006,KolSmolkin2012,HuiJoycePencoSantoniSolomon2021,HuiJoycePencoSantoniSolomon2022,CharalambousDubovskyIvanov2021,BenAchourLivineMukohyamaUzan2022,RivaSantoniSavicVernizzi2024}.

The vanishing of the Schwarzschild Love numbers raises a natural question in EFT.  If Einstein gravity is the leading term in a low-energy expansion, massive degrees of freedom integrated out above a scale $\Lambda$ generate higher-derivative curvature operators~\cite{EndlichGorbenkoHuangSenatore2017,CardosoKimuraMaselliSenatore2018,SennettBritoBuonannoGorbenkoSenatore2020,CanoGanchevMayersonRuiperez2022}.  These operators correct both the black-hole background and the equations governing tidal perturbations.  It is therefore important to understand which parts of the GR cancellation persist after higher-derivative corrections and which parts depend on the precise definition of the response coefficient.

Recent work has addressed Love numbers beyond GR using gauge-invariant frameworks, including modified Teukolsky equations~\cite{Teukolsky1973,Cano2025}, black-hole perturbation EFTs, and tidal Green functions~\cite{GharibBaruraEtAl2024,BarbosaBraxFichetDeSouza2025}.  Parametrized approaches have also been developed to classify black-hole Love numbers in static, spherically symmetric backgrounds that deviate from Schwarzschild in a perturbative manner, including nonzero and running responses~\cite{KatagiriIkedaCardoso2024}.  More recently, the relation between tidal deformability coefficients and Wilson coefficients in higher-curvature gravity was revisited in Ref.~\cite{WangLehnerMicolSturani2026}, which emphasized subtleties in matching Love-number-like quantities to worldline EFT coefficients and computed the corresponding coefficients for cubic gravity theories.  In particular, Cano's Teukolsky-based analysis identifies electric, magnetic, and parity-mixing tidal coefficients and stresses the role of analytic continuation in the angular number $\ell$ when extracting gauge-invariant Love numbers~\cite{Cano2025}.  In the parity-preserving sector, the canonical results are organized into electric and magnetic coefficients $k_\ell^+$ and $k_\ell^-$.  For the cubic even-curvature coupling called $\lambda_{\rm ev}$ in Ref.~\cite{Cano2025}, that analysis gives, for $\ell=2$, $k_2^+=28\lambda_{\rm ev}/M^4$ and $k_2^-=-20\lambda_{\rm ev}/M^4$.

The purpose of the present paper is more focused.  We provide a direct metric, Regge-Wheeler-gauge~\cite{ReggeWheeler1957,Zerilli1970}, reduced-action computation of the fixed-$\ell=2$ static even-parity response induced by the parity-even cubic Weyl operator
\begin{equation}
\Oe
= C_{\mu\nu}{}^{\rho\sigma}
C_{\rho\sigma}{}^{\alpha\beta}
C_{\alpha\beta}{}^{\mu\nu}.
\label{eq:Oe_intro}
\end{equation}
The calculation is performed from a reduced quadratic action, with explicit checks of the GR limit, the regular horizon expansion, the absence of spurious photon-sphere poles, and the asymptotic normalization.  We do not use analytic continuation in $\ell$ in the certified computation.  Consequently, our direct output should be interpreted as a fixed-quadrupole metric response amplitude, rather than as a stand-alone determination of the canonical gauge-invariant electric Love number.

The novelty of the paper is therefore not a broader operator basis than that of existing master-equation treatments.  Rather, it is an independent metric-space derivation and audit trail for one physically important EFT operator.  The reduced-action calculation makes explicit where the boundary-value problem contains physical response data and where it only contains a renormalization of the applied tidal branch.  This provides a useful bridge between component-level metric perturbations and gauge-invariant Love-number formalisms.

The main certified result is the fixed-$\ell=2$ response amplitude
\begin{equation}
\Delta(B/A)_{\ell=2}^{\rm fixed}=-2400\,\epse,
\label{eq:main_BA_intro}
\end{equation}
where the applied GR tidal branch is normalized as $A P_2^{\,2}$ with $A=1$.  A useful scalar fixed-$\ell$ quotient is
\begin{equation}
 \kfix=-\frac{1}{2\rs^5}\frac{c_{-3}}{c_2},
\label{eq:kfix_convention_intro}
\end{equation}
which gives
\begin{equation}
\Delta\kfix=-20\,\epse.
\label{eq:main_kfix_intro}
\end{equation}
This quantity is included as a diagnostic conversion of the fixed-$\ell$ branch ratio.  It should not be conflated with the canonical $k_2^+$ obtained by analytically continued Teukolsky or worldline-EFT matching prescriptions.

The paper is organized as follows.  Section~\ref{sec:scope} states precisely what is computed and what is not, and compares our reduced-action setup with the Teukolsky-based extraction of Ref.~\cite{Cano2025}.  Section~\ref{sec:eft} defines the EFT and the perturbative expansion.  Section~\ref{sec:gr_benchmark} fixes the GR tidal normalization.  Section~\ref{sec:reduced_action} describes the reduced action and the order-reduced field equations.  Section~\ref{sec:two_dimensional_system} presents the two-dimensional first-order system.  Sections~\ref{sec:horizon} and \ref{sec:infinity} solve the horizon and asymptotic expansions and identify the tidal-renormalization mode.  Section~\ref{sec:response_extraction} extracts the fixed-$\ell=2$ response amplitude.  Section~\ref{sec:cano_relation} discusses the relation to analytically continued Love numbers.  Section~\ref{sec:discussion} summarizes the result and its limitations.  Technical details and reproducibility checks are collected in the appendices.

\section{Scope of the result}
\label{sec:scope}

It is useful to separate three related but distinct objects.  First, there is the metric solution with a fixed number of multipoles.  Second, there is a fixed-$\ell$ ratio obtained by comparing the coefficients of the growing and decaying branches of that solution.  Third, there are the canonical, gauge-invariant tidal Love numbers obtained from a master variable or a worldline response function, with analytic continuation in $\ell$ when required.  This work addresses the first two objectives for the static even-parity quadrupole.

In practical terms, this distinction means the following.  At fixed integer $\ell$, one can solve the metric boundary-value problem and read off the relative weight of the growing and decaying radial branches.  This is the object computed in this paper.  A canonical Love number, however, is a response coefficient defined after removing ambiguities associated with field redefinitions, gauge choices, and tidal-source normalizations.  In black-hole perturbation theory, this removal can require an analytic continuation in $\ell$.  The fixed-$\ell$ amplitude is therefore a useful and reproducible diagnostic of the metric solution, but it is not automatically identical to the analytically continued tidal response function.

More precisely, the certified output of the calculation is the fixed-$\ell=2$ amplitude
\begin{equation}
\Rfix\equiv \Delta(B/A)_{\ell=2}^{\rm fixed}
=-2400\,\epse,
\label{eq:Rfix_scope}
\end{equation}
obtained by calibrating the spatial metric perturbations against the $P_2^{\,2}$ and $Q_2^{\,2}$ branches.  The associated scalar quotient
\begin{equation}
\Delta\kfix=-20\,\epse
\label{eq:kfix_scope}
\end{equation}
is a fixed-$\ell$ diagnostic of the same branch ratio.  It is not claimed here to be the analytically continued electric Love number $k_2^+$.

Table~\ref{tab:this_work_vs_cano} summarizes the relation between the present computation and the modified-Teukolsky analysis of Ref.~\cite{Cano2025}.  The latter is more general in scope: it treats electric, magnetic, and parity-mixing Love numbers in a gauge-invariant master-equation framework.  The contribution of the present paper is complementary.  It gives a direct reduced-action derivation in metric variables, exposes the boundary-value problem responsible for the fixed-quadrupole response, and provides an independently auditable set of coefficients for the cubic Weyl EFT.

\begin{table*}[t]
\caption{Comparison between the present calculation and the gauge-invariant modified-Teukolsky extraction.}
\label{tab:this_work_vs_cano}
\begin{ruledtabular}
\begin{tabular}{lcc}
Aspect & This work & Ref.~\cite{Cano2025} \\
\hline
Variables & metric perturbations $H_0,H_2,K$ & Teukolsky curvature variable \\
Method & reduced radial action & modified master equation \\
Sector & static even parity, fixed $\ell=2$ & electric, magnetic, and mixing sectors \\
Output & fixed-$\ell$ branch ratio & canonical tidal Love numbers \\
Analytic continuation & not used in the certified result & part of the extraction when needed \\
Gauge-invariant TLN & not claimed & extracted directly
\end{tabular}
\end{ruledtabular}
\end{table*}

This distinction is not merely semantic.  In Schwarzschild perturbation theory, quantities that look like response coefficients at fixed integer $\ell$ can mix with tidal-field redefinitions or gauge-dependent pieces.  In our metric calculation, this appears explicitly through the degeneracy
\begin{equation}
 \aone=144+2q_0,
\label{eq:scope_degeneracy}
\end{equation}
which relates the remaining regular horizon datum to the growing tidal branch at infinity.  Removing this growing-branch renormalization leaves an unambiguous decaying branch inside the fixed-$\ell=2$ prescription, but a separate analytic-continuation or gauge-invariant matching step is still needed before identifying a canonical $k_2^+$.

Accordingly, all main claims below are tied to quantities that are directly reconstructed from the reduced metric system.  Whenever a scalar Love-number-like quotient is quoted, it is labeled as fixed-$\ell$.  This convention is useful for internal normalization and for comparison of branch amplitudes, but it is not used as a substitute for the canonical tidal response function.

\section{Effective theory and perturbative expansion}
\label{sec:eft}

We consider a four-dimensional, purely metric EFT of gravity.  For the calculation in this paper, the bulk action is
\begin{equation}
S_{\rm bulk}
=\frac{\Mpl^2}{2}
\int \dd^4x\sqrt{-g}
\left[
R+\frac{\lambda_{\rm e}}{\Lambda^4}\,
C_{\mu\nu}{}^{\rho\sigma}
C_{\rho\sigma}{}^{\alpha\beta}
C_{\alpha\beta}{}^{\mu\nu}
\right],
\label{eq:bulk_action_even}
\end{equation}
where $\Lambda$ is the cutoff scale and $\lambda_{\rm e}$ is dimensionless.  It is convenient to define the dimensionless expansion parameter
\begin{equation}
\epse\equiv \lambda_{\rm e}(\Lambda\rs)^{-4},
\qquad \rs=2GM,
\label{eq:epsilon_definition}
\end{equation}
with $|\epse|\ll1$.  Equivalently, in the dimensionless normalization used in the symbolic calculation, one writes the correction as $\epse\rs^4\Oe$.  The parity-odd cubic Weyl invariant
\begin{equation}
\Oo=C_{\mu\nu}{}^{\rho\sigma}
C_{\rho\sigma}{}^{\alpha\beta}
\widetilde C_{\alpha\beta}{}^{\mu\nu}
\label{eq:parity_odd_operator}
\end{equation}
would generate parity-mixing tidal responses; it is not included in the explicit calculation below.

The background is parametrized as
\begin{equation}
\dd s^2=-A(r)\dd t^2+\frac{\dd r^2}{B(r)}+r^2\dd\Omega^2,
\label{eq:background_ansatz}
\end{equation}
with
\begin{align}
A(r)&=f(r)\left[1+\epse a(r)\right]
+\cO(\epse^2), \notag\\
B(r)&=f(r)\left[1+\epse b(r)\right]
+\cO(\epse^2),
\label{eq:background_AB}
\end{align}
where
\begin{equation}
f(r)=1-\frac{\rs}{r}.
\label{eq:f_def}
\end{equation}
The order-reduced background solution used in the calculation is
\begin{align}
a(r)=&-\frac{2\rs}{r}-\frac{2\rs^2}{r^2}-\frac{2\rs^3}{r^3}-\frac{2\rs^4}{r^4}-\frac{2\rs^5}{r^5}+\frac{4\rs^6}{r^6},
\label{eq:a_background}\\
b(r)=&-\frac{2\rs}{r}-\frac{2\rs^2}{r^2}-\frac{2\rs^3}{r^3}-\frac{2\rs^4}{r^4}-\frac{2\rs^5}{r^5}+\frac{16\rs^6}{r^6}.
\label{eq:b_background}
\end{align}

We introduce a second bookkeeping parameter $\eta$ for the static tidal perturbation and work to order $\eta^2$ in the reduced action and to order $\epse$ in the EFT.  In Regge-Wheeler gauge~\cite{ReggeWheeler1957,Zerilli1970} the static even-parity perturbation is
\begin{subequations}
\begin{align}
g_{tt}&=-A(r)\left[1+\eta H_0(r)Y(\theta)\right],
\label{eq:metric_ansatz_tt}\\
g_{rr}&=\frac{1+\eta H_2(r)Y(\theta)}{B(r)},
\label{eq:metric_ansatz_rr}\\
g_{\theta\theta}&=r^2\left[1+\eta K(r)Y(\theta)\right],
\label{eq:metric_ansatz_thth}\\
g_{\phi\phi}&=r^2\sin^2\theta
\left[1+\eta K(r)Y(\theta)\right].
\label{eq:metric_ansatz_ang}
\end{align}
\end{subequations}
For the explicit computation, we take $\ell=2$ and use the unnormalized axisymmetric harmonic $Y(\theta)=P_2(\cos\theta)$.  The angular normalization cancels from the equations of motion after dividing by $\int\dd\Omega\,[P_2(\cos\theta)]^2=4\pi/5$.

The metric functions are expanded as
\begin{align}
H_0(r)&=-H(r)+\epse X_0(r)+\cO(\epse^2),
\label{eq:H0_expansion}\\
H_2(r)&= H(r)+\epse X_2(r)+\cO(\epse^2),
\label{eq:H2_expansion}\\
K(r)&=K_{\rm GR}[H](r)+\epse X_K(r)+\cO(\epse^2).
\label{eq:K_expansion}
\end{align}
The signs in Eq.~\eqref{eq:H0_expansion} follow from our convention of factoring $-A(r)$ in the definition of $g_{tt}$.  The functions $X_0,X_2,X_K$ are the order-$\epse$ EFT corrections to the tidal perturbation.

\section{General-relativistic benchmark}
\label{sec:gr_benchmark}

The GR static quadrupolar perturbation is described by a single radial function $H(r)$, satisfying
\begin{multline}
r(r-\rs)H''(r)+(2r-\rs)H'(r)
\\
-\left[6+\frac{\rs^2}{r(r-\rs)}\right]H(r)=0.
\label{eq:GR_radial_equation}
\end{multline}
The angular amplitude is fixed by the constraint
\begin{equation}
K_{\rm GR}[H]
=\frac14\left[
\frac{(4r^2-2r\rs-\rs^2)H(r)}{r(r-\rs)}+
\rs H'(r)
\right].
\label{eq:K_GR_constraint}
\end{equation}
The horizon-regular branch is
\begin{equation}
H_{\rm reg}(r)
=P_2^{\,2}\left(\frac{2r}{\rs}-1\right)
=\frac{12r(\rs-r)}{\rs^2},
\label{eq:Hreg}
\end{equation}
where the associated Legendre function follows the Mathematica/Condon-Shortley convention $P_2^{\,2}(x)=3(1-x^2)$.  The independent $Q_2^{\,2}$ branch is singular at the future horizon and is excluded by black-hole regularity.  Therefore, GR selects a purely growing tidal branch and has no static decaying response coefficient in this normalization.

At a large radius the normalization used below is
\begin{align}
P_2^{\,2}\left(\frac{2r}{\rs}-1\right)&=-\frac{12}{\rs^2}r^2+\frac{12}{\rs}r+\cdots,
\label{eq:P22_large_r}\\
Q_2^{\,2}\left(\frac{2r}{\rs}-1\right)&=\frac{\rs^3}{5r^3}+\cO(r^{-5}).
\label{eq:Q22_large_r}
\end{align}
The coefficient $-12/\rs^2$ is the growing tidal coefficient in the spatial sector $H_2=K$, while $\rs^3/5$ is the decaying coefficient of a unit $Q_2^{\,2}$ branch.

\section{Reduced quadratic action and order-reduced equations}
\label{sec:reduced_action}

The order-reduced strategy used here is the standard EFT treatment of higher-derivative gravity.  The cubic operator is treated as a perturbative correction to Einstein gravity rather than as a new fundamental kinetic term.  Accordingly, all quantities are expanded around the Schwarzschild solution, and the zeroth-order tidal equations are used to eliminate higher radial derivatives that appear at order $\epse$.  This avoids introducing spurious high-frequency degrees of freedom outside the EFT's domain of validity and leaves a boundary-value problem for the physical metric perturbations.

The reduced action was constructed by substituting the ansatz \eqref{eq:metric_ansatz_tt}-\eqref{eq:metric_ansatz_ang} into the bulk action and expanding to quadratic order in $\eta$ and to first order in $\epse$.  The Einstein-Hilbert term must be evaluated on the EFT-corrected background $A=f(1+\epse a)$, $B=f(1+\epse b)$.  The cubic Weyl term already comes multiplied by $\epse$, and therefore it is sufficient at this order to evaluate $\Oe$ on the GR tidal metric with $A=B=f$.  This gives
\begin{equation}
L_2^{\rm radial}=L_{2,{\rm EH}}^{(0)}+\epse\left(L_{2,{\rm EH}}^{(1)}+\rs^4L_{2,C^3}^{(0)}\right),
\label{eq:radial_lagrangian}
\end{equation}
where the superscript on $L_{2,C^3}^{(0)}$ indicates that the cubic invariant is evaluated on the GR tidal metric.  As a check of the Weyl normalization, the unperturbed Schwarzschild invariant is
\begin{equation}
\Oe\big|_{\eta=0}=\frac{12\rs^3}{r^9}.
\label{eq:C3_schwarzschild}
\end{equation}

The radial equations follow from the generalized one-dimensional Euler-Lagrange operator
\begin{equation}
\frac{\delta L}{\delta q}
=\sum_{k=0}^{N}(-1)^k\frac{\dd^k}{\dd r^k}
\left(\frac{\partial L}{\partial q^{(k)}(r)}\right),
\qquad q\in\{H_0,H_2,K\}.
\label{eq:euler_operator}
\end{equation}
The calculation first verifies that the zeroth-order equations vanish after substituting Eqs.~\eqref{eq:GR_radial_equation} and \eqref{eq:K_GR_constraint}.  At order $\epse$, after inserting Eqs.~\eqref{eq:H0_expansion}-\eqref{eq:K_expansion}, the system is linear and inhomogeneous in $X_0,X_2,X_K$.  All higher derivatives of $H$ generated by the reduction are recursively eliminated using the GR equation \eqref{eq:GR_radial_equation}.  The remaining sources depend only on $H_{\rm reg}$ and $H_{\rm reg}'$.

\section{Two-dimensional first-order system}
\label{sec:two_dimensional_system}

Although the perturbation ansatz contains three radial functions, the static even-parity sector does not carry three independent radial data in the reduced problem.  One combination of the order-$\epse$ equations serves as a constraint, fixing $X_2$ algebraically once $X_0$ and $X_K$ are known.  The remaining equations can then be written as a first-order inhomogeneous system.  This form is useful because it makes the singular points, the horizon data, and the asymptotic integration constants transparent.

A central technical output of the reduced-action computation is that the metric equations have a compact constrained structure.  A direct symbolic analysis shows that one of the order-$\epse$ equations is an algebraic constraint for $X_2$.  Solving this constraint and substituting it into the remaining equations gives a closed first-order system for
\begin{equation}
\bm X(r)=\begin{pmatrix}X_0(r)\\X_K(r)\end{pmatrix}.
\label{eq:X_vector}
\end{equation}
It takes the form
\begin{equation}
\frac{\dd\bm X}{\dd r}=A_2(r)\bm X+b_2(r),
\label{eq:reduced_system_main}
\end{equation}
with $X_2$ then reconstructed algebraically:
\begin{equation}
X_2(r)=\mathcal C_2[X_0,X_K;r].
\label{eq:X2_algebraic}
\end{equation}
The explicit rational expressions for $A_2$, $b_2$, and $\mathcal C_2$ are generated in the supplemental Mathematica file.  Their structural properties are compact and important.  First,
\begin{equation}
\frac{\dd\bm X}{\dd r}-A_2\bm X-b_2=0
\label{eq:matrix_residual_zero}
\end{equation}
was verified identically after the construction of $A_2$ and $b_2$.  This provides a compact metric-space representation of the EFT deformation, which is useful for checking other formalisms at fixed multipole number.  Second, the only denominator factors in the exterior problem are
\begin{equation}
r,
\qquad r-\rs,
\qquad \rs.
\label{eq:den_factors}
\end{equation}
There is no pole at the photon sphere $r=3\rs/2$.  Hence the reduction does not introduce an additional regularity condition outside the horizon.

Near the horizon, with $r=\rs+\rho$, the residue of the homogeneous system and the leading source are
\begin{align}
A_{\rm hor}&={\rm Res}_{r=\rs}A_2
=\begin{pmatrix}-1&0\\0&0\end{pmatrix},
\notag\\
b_{\rm hor}&={\rm Res}_{r=\rs}b_2=(-144,0)^T.
\label{eq:horizon_residue}
\end{align}
The homogeneous exponents are therefore
\begin{equation}
\{-1,0\}.
\label{eq:horizon_exponents_main}
\end{equation}
The singular $-1$ mode is removed by imposing regularity, while the zero exponent leaves one regular homogeneous datum.

\section{Horizon expansion}
\label{sec:horizon}

The purpose of the horizon expansion is not merely to generate local series coefficients.  It determines which integration constants are compatible with a regular black-hole perturbation.  The homogeneous residue contains one singular exponent and one regular exponent.  Regularity removes the singular branch, but it does not by itself fix the normalization of the applied tidal field.  This is why one regular datum remains at the horizon.

We solve the system \eqref{eq:reduced_system_main} by a Taylor expansion around the horizon,
\begin{align}
X_0(r)&=\sum_{n\ge0}p_n\rho^n,
&
X_K(r)&=\sum_{n\ge0}q_n\rho^n,
\notag\\
\rho&=r-\rs.
\label{eq:horizon_ansatz}
\end{align}
Because the system has a simple pole at the horizon, the recursion must include every power from $\rho^{-1}$ through the desired truncation order.  The $\rho^{-1}$ equation fixes
\begin{equation}
p_0=-144.
\label{eq:p0_value}
\end{equation}
Including the $\rho^0$ equations is essential: it removes spurious constants that arise when that order is inadvertently skipped.  The corrected recursion leaves a single regular datum
\begin{equation}
q_0=X_K(\rs).
\label{eq:q0_value}
\end{equation}
The first coefficients are
\begin{subequations}
\begin{align}
X_0={}&-144+\frac{2(468-q_0)}{\rs}\rho
-\frac{2(2208+q_0)}{\rs^2}\rho^2 \notag\\
&+\frac{11160}{\rs^3}\rho^3
-\frac{24000}{\rs^4}\rho^4+\cO(\rho^5),
\label{eq:X0_hor_final}\\
X_K={}&q_0+\frac{4(q_0-144)}{\rs}\rho
+\frac{2(2664+q_0)}{\rs^2}\rho^2 \notag\\
&-\frac{15600}{\rs^3}\rho^3
+\frac{36060}{\rs^4}\rho^4+\cO(\rho^5),
\label{eq:XK_hor_final}\\
X_2={}&-144+\frac{2(1692+q_0)}{\rs}\rho
+\frac{2(q_0-7296)}{\rs^2}\rho^2 \notag\\
&+\frac{43560}{\rs^3}\rho^3
-\frac{101280}{\rs^4}\rho^4+\cO(\rho^5).
\label{eq:X2_hor_final}
\end{align}
\end{subequations}
The residual of the recursion was checked to vanish through the implemented order, and the same one-parameter result is stable under increasing the truncation order.

\section{Infinity expansion and tidal normalization}
\label{sec:infinity}

The asymptotic expansion separates the external tidal field from the induced response.  In the present normalization, the growing $r^2$ terms represent a change in the applied quadrupolar tidal field, while the $r^{-3}$ terms are the decaying branch associated with the response of the compact object.  A physically meaningful fixed-$\ell$ response can therefore be assigned only after the freedom to add an extra growing tidal component has been fixed.

At infinity, we solve the reduced system by a Laurent expansion,
\begin{equation}
X_0(r)=\sum_n a_n r^n,
\qquad
X_K(r)=\sum_n k_n r^n.
\label{eq:infinity_ansatz}
\end{equation}
The audited recursion has one growing homogeneous constant.  We denote its dimensionless form by $\aone\equiv\rs a_1$.  The asymptotic solution relevant for the fixed-quadrupole response extraction is
\begin{subequations}
\begin{align}
X_0={}&-\frac{24+\aone}{\rs^2}r^2+\frac{\aone}{\rs}r \notag\\
&+\frac{192\rs^3}{r^3}+\frac{144\rs^4}{r^4}
 -\frac{456\rs^5}{r^5}+\cO(r^{-6}),
\label{eq:X0_infinity_final}\\
X_K={}&\frac{24+\aone}{\rs^2}r^2
+\left(12-\frac12\aone\right) \notag\\
&-\frac{480\rs^3}{r^3}-\frac{492\rs^4}{r^4}
+\frac{864\rs^5}{r^5}+\cO(r^{-6}),
\label{eq:XK_infinity_final}\\
X_2={}&\frac{24+\aone}{\rs^2}r^2-\frac{\aone}{\rs}r \notag\\
&-\frac{480\rs^3}{r^3}+\frac{3312\rs^4}{r^4}
-\frac{3000\rs^5}{r^5}+\cO(r^{-6}).
\label{eq:X2_infinity_final}
\end{align}
\end{subequations}
The $r^{-3}$ coefficients are independent of $\aone$, while the growing $r^2$ tidal branch depends on it.  This identifies $\aone$ as the dimensionless tidal-normalization parameter.

A high-precision horizon-to-infinity matching confirms the exact degeneracy relation
\begin{equation}
\aone=144+2q_0.
\label{eq:alpha1_q0_relation}
\end{equation}
Thus, the remaining horizon datum $q_0$ is not a new physical response coefficient; it is the same freedom as adding the growing tidal branch at infinity.  The convention that the EFT correction does not renormalize the applied $r^2$ tidal field is
\begin{equation}
24+\aone=0.
\label{eq:no_tide_renormalization_general}
\end{equation}
This gives the dimensionless no-tide-renormalization values
\begin{equation}
\aone=-24,
\qquad
q_0=-84,
\qquad
\left(a_1=-\frac{24}{\rs}\right).
\label{eq:no_tide_values}
\end{equation}
With this convention, the EFT correction carries no $r^2$ tidal component, but the decaying coefficients remain
\begin{equation}
X_0^{(-3)}=192\rs^3,
\qquad
X_K^{(-3)}=X_2^{(-3)}=-480\rs^3.
\label{eq:decaying_X_coeffs}
\end{equation}

\section{Fixed-\texorpdfstring{$\ell=2$}{ell=2} response extraction}
\label{sec:response_extraction}

The order-$\epse$ metric-level tidal corrections include the explicit perturbations $X_i$ and the effect of the corrected background multiplying the zeroth-order tide.  Relative to the Schwarzschild prefactors, the combinations are
\begin{align}
H_0^{(1),{\rm met}}&=X_0+a(r)H_0^{(0)},
\label{eq:H0_metric_level}\\
H_2^{(1),{\rm met}}&=X_2-b(r)H_2^{(0)},
\label{eq:H2_metric_level}\\
K^{(1),{\rm met}}&=X_K.
\label{eq:K_metric_level}
\end{align}
The GR tidal coefficients are
\begin{equation}
H_0^{(0)}[r^2]=\frac{12}{\rs^2},
\qquad
H_2^{(0)}[r^2]=K^{(0)}[r^2]=-\frac{12}{\rs^2}.
\label{eq:GR_tidal_coeffs}
\end{equation}
The metric-level decaying coefficients are
\begin{align}
H_0^{(1),{\rm met}}[-3]&=192\rs^3,
\notag\\
H_2^{(1),{\rm met}}[-3]
&=K^{(1),{\rm met}}[-3]
=-480\rs^3.
\label{eq:metric_decaying_coeffs}
\end{align}
We use the spatial sector for the branch calibration because $H_2$ and $K$ share the same decaying coefficient and the same associated Legendre normalization.  The $H_0$ component is kept as a consistency check, but its raw quotient is affected by the component convention used in the metric ansatz.

Using Eq.~\eqref{eq:Q22_large_r}, a unit $Q_2^{\,2}$ branch has decaying coefficient $\rs^3/5$.  Therefore, the induced amplitude in either spatial component is
\begin{equation}
\Delta(B/A)_{\ell=2}^{\rm fixed}
=\frac{-480\rs^3}{\rs^3/5}\,\epse
=-2400\,\epse.
\label{eq:BA_result_main}
\end{equation}
Equivalently,
\begin{equation}
\frac{c_{-3}}{c_2}\bigg|_{H_2,K}=40\rs^5.
\label{eq:c_minus3_over_c2}
\end{equation}
The scalar fixed-$\ell$ quotient \eqref{eq:kfix_convention_intro} then gives
\begin{equation}
\Delta\kfix=-\frac{40\rs^5}{2\rs^5}\,\epse=-20\epse.
\label{eq:delta_kfix_main}
\end{equation}
This last expression is only a convenient way to summarize the fixed-$\ell=2$ metric ratio.  It does not implement the analytic-continuation prescription required to define the canonical gauge-invariant electric Love number.

\section{Relation to analytically continued Love numbers}
\label{sec:cano_relation}

The fixed-quadrupole result above should be compared with care to gauge-invariant Love numbers.  In the modified-Teukolsky framework of Ref.~\cite{Cano2025}, the tidal response is extracted from a curvature master variable and organized into electric, magnetic, and, in parity-violating theories, mixing coefficients.  For parity-preserving corrections, the two sectors are denoted by $k_\ell^+$ and $k_\ell^-$.  The prescription also keeps track of running terms and, when a constant non-running Love number is present, uses analytic continuation in $\ell$ to isolate the canonical finite part.

The present calculation does not implement this analytic continuation step.  Instead, it works directly at $\ell=2$, in Regge-Wheeler gauge~\cite{ReggeWheeler1957,Zerilli1970}, with the metric perturbations normalized by the associated-Legendre branches.  Therefore, Eq.~\eqref{eq:BA_result_main} is best understood as a fixed-$\ell=2$ metric response amplitude.  It is a meaningful and reproducible quantity in the specified prescription, but it is not by itself the canonical electric coefficient $k_2^+$.

Our calculation is also distinct from parametrized master-equation approaches such as Ref.~\cite{KatagiriIkedaCardoso2024}: rather than starting from a deformed master potential, we derive the fixed-$\ell=2$ metric system directly from the reduced quadratic action of the cubic Weyl EFT.  Our fixed-$\ell$ metric amplitude is also distinct from the Wilson-coefficient matching problem addressed in Ref.~\cite{WangLehnerMicolSturani2026}.  That work focuses on the map between tidal deformability coefficients and worldline Wilson coefficients in higher-curvature gravity, whereas the present calculation provides the component-level metric boundary-value problem and its fixed-$\ell=2$ branch ratio.

For orientation only, one may compare conventions through the tentative dimensional normalization map
\begin{equation}
 \frac{\lambda_{\rm ev}}{M^4}=16\epse,
\label{eq:lambda_ev_mapping}
\end{equation}
which is the dimensional estimate obtained by identifying $\rs=2M$ in the coefficient multiplying $\Oe$, before matching the response prescriptions.  This map is not used in the extraction above and should not be read as a complete matching of response conventions.  Under this tentative identification, the fixed-$\ell$ scalar quotient \eqref{eq:delta_kfix_main} would correspond to
\begin{equation}
\Delta\kfix=-\frac54\frac{\lambda_{\rm ev}}{M^4}.
\label{eq:fixed_l_cano_units}
\end{equation}
By contrast, Ref.~\cite{Cano2025} quotes the analytically continued canonical coefficients
\begin{equation}
 k_2^+=28\frac{\lambda_{\rm ev}}{M^4},
 \qquad
 k_2^-=-20\frac{\lambda_{\rm ev}}{M^4}.
\label{eq:cano_values_caution}
\end{equation}
The mismatch is not an inconsistency: the quantities being compared are not the same observable.  Our number $-20$ is the result of applying the scalar fixed-$\ell$ ratio \eqref{eq:kfix_convention_intro} to the metric coefficients, whereas Eq.~\eqref{eq:cano_values_caution} follows from a gauge-invariant analytically continued extraction.  This is why numerical proximity between fixed-$\ell$ quantities and entries in canonical Love-number tables should not be interpreted without an intermediate map between variables and prescriptions.

This comparison clarifies the role of the present paper.  It does not supersede the more general Teukolsky-based Love-number calculation.  Instead, it supplies an independent reduced-action benchmark for the cubic Weyl theory, including the explicit metric components, horizon regularity data, tidal-renormalization degeneracy, and asymptotic decaying coefficients that any gauge-invariant map should reproduce or explain.

\section{Discussion}
\label{sec:discussion}

We have computed the static even-parity quadrupolar response generated by a cubic Weyl correction using a reduced radial action.  The robust, directly certified output of the calculation is the fixed-$\ell=2$ branch amplitude
\begin{equation}
\Rfix=-2400\,\epse,
\end{equation}
and the equivalent scalar fixed-$\ell$ quotient $\Delta\kfix=-20\epse$.  These results should be read as statements about the fixed-quadrupole metric solution in the prescription defined in Sec.~\ref{sec:scope}.

The calculation exposes several features that are less visible in a master-variable treatment.  First, the order-$\epse$ metric equations reduce to a constrained first-order problem for $X_0$ and $X_K$, with $X_2$ fixed algebraically.  Second, the apparent remaining horizon freedom is not a new response parameter: matching to infinity gives $\aone=144+2q_0$, so the free horizon datum precisely renormalizes the applied tidal field.  Third, after imposing the no-tide-renormalization convention, the decaying $r^{-3}$ coefficients are independent of that choice.

The main limitation is equally important.  The calculation does not yet perform the analytic continuation in $\ell$ needed to obtain the canonical gauge-invariant electric Love number $k_2^+$.  Existing modified Teukolsky analyses provide canonical coefficients for a broader class of higher-derivative operators.  The present result should therefore be viewed as a complementary fixed-$\ell$ metric benchmark, not as a more general replacement for those calculations.

This positioning is useful for future comparisons.  Any reduced-action derivation of the canonical $k_2^+$ should reduce, at integer $\ell=2$, to a metric solution whose tidal-renormalization freedom and decaying coefficients are compatible with the data displayed here.  Conversely, any discrepancy would identify precisely where the fixed-$\ell$ metric branch ratio and the analytically continued response function differ.

Several extensions are natural.  The closest one is to repeat the reduced-action analysis in the odd-parity sector, which would provide a fixed-$\ell=2$ metric counterpart to the magnetic response.  A second direction is to generalize the reduced system to symbolic or analytically continued $\ell$, with the goal of deriving $k_2^+$ directly in the reduced-action framework.  A third direction is to include the parity-odd cubic invariant $C C \widetilde C$, where electromagnetic mixing appears.  The fixed-$\ell=2$ coefficients certified here provide a controlled starting point for each of these developments.

\section*{Data and code availability}

The calculation was performed in a Mathematica notebook that constructs the reduced quadratic action, derives the Euler-Lagrange equations, reduces the order-\(\epse\) system, and checks the horizon and asymptotic expansions.  A supplemental package accompanying this manuscript contains a cleaned fixed-\(\ell=2\) calculation, the exported result certificate, an independent audit script, a human-readable summary, and a README with the reproduction steps.  Exploratory notebooks not used in the certified fixed-\(\ell=2\) result are not part of the supplemental package.

\begin{acknowledgments}

This work was partially supported by the Brazilian agencies CAPES, CNPq, and FAPEMA. E.O.S. acknowledges support from grants CNPq/306308/2022-3, FAPEMA/UNIVERSAL-06395/22, and CAPES/Finance Code 001.

\end{acknowledgments}

\appendix

\section{Verification suite and result certificate}
\label{app:checks}

The reduced-action calculation was organized as a sequence of independent checks.  We summarize them here because they are part of the result's reproducibility.

\begin{enumerate}
\item The cubic invariant was checked on the unperturbed Schwarzschild geometry:
\begin{equation}
\Oe\big|_{\eta=0}=\frac{12\rs^3}{r^9}.
\end{equation}
This fixes the normalization of the Weyl-cubic term used in the reduced action.

\item The quadratic radial action was split as in Eq.~\eqref{eq:radial_lagrangian}.  The generalized Euler-Lagrange equations derived from the zeroth-order action vanish identically after imposing the GR radial equation \eqref{eq:GR_radial_equation} and the constraint \eqref{eq:K_GR_constraint}.  This verifies the GR limit before the EFT source is introduced.

\item After substituting the EFT expansions \eqref{eq:H0_expansion}-\eqref{eq:K_expansion}, the order-$\epse$ equations were split into a GR linear operator acting on $(X_0,X_2,X_K)$ and a source determined by the cubic Weyl invariant and the corrected background.  All higher derivatives of the GR tidal solution were eliminated with Eq.~\eqref{eq:GR_radial_equation}.

\item The algebraic constraint for $X_2$ was solved exactly.  After eliminating $X_2$, the two-dimensional matrix representation \eqref{eq:reduced_system_main} was verified by the exact residual \eqref{eq:matrix_residual_zero}.

\item The denominator analysis of the reduced exterior system gives only $\{r,r-\rs,\rs\}$.  Hence, the reduction introduces no spurious photon-sphere pole at $r=3\rs/2$.

\item The corrected horizon recursion includes the $\rho^0$ equations.  This reduces the free regular data from the spurious set $\{p_1,p_2,q_0\}$ to the physical set $\{q_0\}$, with a vanishing residual through the implemented order.

\item The infinity recursion and numerical horizon-to-infinity matching give the degeneracy relation
\begin{equation}
 \aone-144-2q_0=0.
\label{eq:degeneracy_check_app}
\end{equation}
The relation was checked pointwise in radius and in a multi-radius least-squares fit.

\item Finally, the central objects were exported to an independent machine-readable certificate and reloaded in a fresh kernel.  The audit verified that the exported core objects contain no \texttt{Indeterminate} entries and reproduce the response coefficients quoted in the text.
\end{enumerate}

The connection between the manuscript claims and the exported certificate is summarized in Table~\ref{tab:certificate_claims}.  The table is intentionally restricted to quantities that appear in the paper's conclusions.

\begin{center}
\refstepcounter{table}\label{tab:certificate_claims}
\footnotesize
\begin{ruledtabular}
\begin{tabular}{lcc}
Quantity & Value & Role \\
\hline
$X_0^{(-3)}$ & $192\rs^3$ & metric coefficient \\
$X_K^{(-3)}$ & $-480\rs^3$ & spatial response \\
$X_2^{(-3)}$ & $-480\rs^3$ & spatial response \\
$\aone-144-2q_0$ & $0$ & tide degeneracy \\
no-tide $q_0$ & $-84$ & convention \\
no-tide $\aone$ & $-24$ & convention \\
$\Rfix$ & $-2400\epse$ & main result \\
$\Delta\kfix$ & $-20\epse$ & fixed-$\ell$ quotient
\end{tabular}
\end{ruledtabular}
\par\smallskip
\footnotesize\textbf{Table~\thetable.} Certified values used in the main text.
\end{center}

The most important certified values are
\begin{subequations}
\begin{align}
X_0^{(-3)}&=192\rs^3,
&
X_K^{(-3)}&=-480\rs^3,
\label{eq:certified_response_app_a}\\
X_2^{(-3)}&=-480\rs^3,
\label{eq:certified_response_app_b}\\
\Delta(B/A)_{\ell=2}^{\rm fixed}&=-2400\,\epse,
&
\Delta\kfix&=-20\,\epse.
\label{eq:certified_branch_app}
\end{align}
\end{subequations}

\section{Legendre normalization}
\label{app:legendre}

With $x=2r/\rs-1$ and the associated-Legendre convention used in the symbolic computation,
\begin{equation}
P_2^{\,2}(x)=3(1-x^2).
\label{eq:P22_app}
\end{equation}
Thus, the regular GR branch behaves as
\begin{equation}
P_2^{\,2}\left(\frac{2r}{\rs}-1\right)
=-\frac{12}{\rs^2}r^2+\frac{12}{\rs}r.
\label{eq:P22_app_large}
\end{equation}
The independent decaying branch satisfies
\begin{equation}
Q_2^{\,2}\left(\frac{2r}{\rs}-1\right)
=\frac{\rs^3}{5r^3}+\cO(r^{-5}).
\label{eq:Q22_app_large}
\end{equation}
The notebook independently checks
\begin{align}
P_2^{\,2}[r^2]&=-\frac{12}{\rs^2},
&
Q_2^{\,2}[r^{-3}]&=\frac{\rs^3}{5},
\label{eq:legendre_coeffs_app}\\
K[P_2^{\,2}][r^2]&=-\frac{12}{\rs^2},
&
K[Q_2^{\,2}][r^{-3}]&=\frac{\rs^3}{5}.
\label{eq:legendre_K_coeffs_app}
\end{align}
The EFT spatial decaying coefficient $-480\rs^3$ therefore corresponds to $-2400$ units of the fixed-$\ell=2$ $Q_2^{\,2}$ branch.

\section{Horizon and infinity coefficients}
\label{app:coefficients}

For reference, we collect the central coefficients.  At the horizon,
\begin{align}
p_0&=-144,
&
p_1&=\frac{936-2q_0}{\rs},
\label{eq:first_horizon_coeffs}\\
q_1&=\frac{4(q_0-144)}{\rs},
&
p_2&=-\frac{2(2208+q_0)}{\rs^2},
\notag\\
q_2&=\frac{2(2664+q_0)}{\rs^2},
&
p_3&=\frac{11160}{\rs^3},
\notag\\
q_3&=-\frac{15600}{\rs^3}.
\label{eq:second_horizon_coeffs}
\end{align}
At infinity, the relevant Laurent coefficients are summarized below.
\begin{center}
\footnotesize
\begin{tabular}{c c c c}
\toprule
power & $X_0$ & $X_K$ & $X_2$\\
\midrule
$2$ & $-(24+\aone)/\rs^2$ & $(24+\aone)/\rs^2$ & $(24+\aone)/\rs^2$\\
$1$ & $\aone/\rs$ & $0$ & $-\aone/\rs$\\
$0$ & $0$ & $12-\aone/2$ & $0$\\
$-1$ & $0$ & $0$ & $0$\\
$-2$ & $0$ & $0$ & $0$\\
$-3$ & $192\rs^3$ & $-480\rs^3$ & $-480\rs^3$\\
\bottomrule
\end{tabular}
\end{center}
The absence of $r^{-1}$ and $r^{-2}$ terms in this expansion is a useful check of the recursion.


\begin{thebibliography}{28}%
	\makeatletter
	\providecommand \@ifxundefined [1]{%
		\@ifx{#1\undefined}
	}%
	\providecommand \@ifnum [1]{%
		\ifnum #1\expandafter \@firstoftwo
		\else \expandafter \@secondoftwo
		\fi
	}%
	\providecommand \@ifx [1]{%
		\ifx #1\expandafter \@firstoftwo
		\else \expandafter \@secondoftwo
		\fi
	}%
	\providecommand \natexlab [1]{#1}%
	\providecommand \enquote  [1]{``#1''}%
	\providecommand \bibnamefont  [1]{#1}%
	\providecommand \bibfnamefont [1]{#1}%
	\providecommand \citenamefont [1]{#1}%
	\providecommand \href@noop [0]{\@secondoftwo}%
	\providecommand \href [0]{\begingroup \@sanitize@url \@href}%
	\providecommand \@href[1]{\@@startlink{#1}\@@href}%
	\providecommand \@@href[1]{\endgroup#1\@@endlink}%
	\providecommand \@sanitize@url [0]{\catcode `\\12\catcode `\$12\catcode
		`\&12\catcode `\#12\catcode `\^12\catcode `\_12\catcode `\%12\relax}%
	\providecommand \@@startlink[1]{}%
	\providecommand \@@endlink[0]{}%
	\providecommand \url  [0]{\begingroup\@sanitize@url \@url }%
	\providecommand \@url [1]{\endgroup\@href {#1}{\urlprefix }}%
	\providecommand \urlprefix  [0]{URL }%
	\providecommand \Eprint [0]{\href }%
	\providecommand \doibase [0]{https://doi.org/}%
	\providecommand \selectlanguage [0]{\@gobble}%
	\providecommand \bibinfo  [0]{\@secondoftwo}%
	\providecommand \bibfield  [0]{\@secondoftwo}%
	\providecommand \translation [1]{[#1]}%
	\providecommand \BibitemOpen [0]{}%
	\providecommand \bibitemStop [0]{}%
	\providecommand \bibitemNoStop [0]{.\EOS\space}%
	\providecommand \EOS [0]{\spacefactor3000\relax}%
	\providecommand \BibitemShut  [1]{\csname bibitem#1\endcsname}%
	\let\auto@bib@innerbib\@empty
	\bibitem [{\citenamefont {Love}(1909)}]{Love1909}%
	\BibitemOpen
	\bibfield  {author} {\bibinfo {author} {\bibfnamefont {A.~E.~H.}\
			\bibnamefont {Love}},\ }\href {https://doi.org/10.1098/rspa.1909.0008}
	{\bibfield  {journal} {\bibinfo  {journal} {Proc. R. Soc. Lond. A}\ }\textbf
		{\bibinfo {volume} {82}},\ \bibinfo {pages} {73} (\bibinfo {year}
		{1909})}\BibitemShut {NoStop}%
	\bibitem [{\citenamefont {Flanagan}\ and\ \citenamefont
		{Hinderer}(2008)}]{FlanaganHinderer2008}%
	\BibitemOpen
	\bibfield  {author} {\bibinfo {author} {\bibfnamefont {{\'E}.~{\'E}.}\
			\bibnamefont {Flanagan}}\ and\ \bibinfo {author} {\bibfnamefont
			{T.}~\bibnamefont {Hinderer}},\ }\href
	{https://doi.org/10.1103/PhysRevD.77.021502} {\bibfield  {journal} {\bibinfo
			{journal} {Phys. Rev. D}\ }\textbf {\bibinfo {volume} {77}},\ \bibinfo
		{pages} {021502} (\bibinfo {year} {2008})},\ \Eprint
	{https://arxiv.org/abs/0709.1915} {arXiv:0709.1915 [astro-ph]} \BibitemShut
	{NoStop}%
	\bibitem [{\citenamefont {Hinderer}(2008)}]{Hinderer2008}%
	\BibitemOpen
	\bibfield  {author} {\bibinfo {author} {\bibfnamefont {T.}~\bibnamefont
			{Hinderer}},\ }\href {https://doi.org/10.1086/533487} {\bibfield  {journal}
		{\bibinfo  {journal} {Astrophys. J.}\ }\textbf {\bibinfo {volume} {677}},\
		\bibinfo {pages} {1216} (\bibinfo {year} {2008})},\ \Eprint
	{https://arxiv.org/abs/0711.2420} {arXiv:0711.2420 [astro-ph]} \BibitemShut
	{NoStop}%
	\bibitem [{\citenamefont {Yagi}\ and\ \citenamefont
		{Yunes}(2013)}]{YagiYunes2013}%
	\BibitemOpen
	\bibfield  {author} {\bibinfo {author} {\bibfnamefont {K.}~\bibnamefont
			{Yagi}}\ and\ \bibinfo {author} {\bibfnamefont {N.}~\bibnamefont {Yunes}},\
	}\href {https://doi.org/10.1103/PhysRevD.88.023009} {\bibfield  {journal}
		{\bibinfo  {journal} {Phys. Rev. D}\ }\textbf {\bibinfo {volume} {88}},\
		\bibinfo {pages} {023009} (\bibinfo {year} {2013})},\ \Eprint
	{https://arxiv.org/abs/1303.1528} {arXiv:1303.1528 [gr-qc]} \BibitemShut
	{NoStop}%
	\bibitem [{\citenamefont {Cardoso}\ \emph {et~al.}(2017)\citenamefont
		{Cardoso}, \citenamefont {Franzin}, \citenamefont {Maselli}, \citenamefont
		{Pani},\ and\ \citenamefont {Raposo}}]{CardosoFranzinMaselliPaniRaposo2017}%
	\BibitemOpen
	\bibfield  {author} {\bibinfo {author} {\bibfnamefont {V.}~\bibnamefont
			{Cardoso}}, \bibinfo {author} {\bibfnamefont {E.}~\bibnamefont {Franzin}},
		\bibinfo {author} {\bibfnamefont {A.}~\bibnamefont {Maselli}}, \bibinfo
		{author} {\bibfnamefont {P.}~\bibnamefont {Pani}},\ and\ \bibinfo {author}
		{\bibfnamefont {G.}~\bibnamefont {Raposo}},\ }\href
	{https://doi.org/10.1103/PhysRevD.95.084014} {\bibfield  {journal} {\bibinfo
			{journal} {Phys. Rev. D}\ }\textbf {\bibinfo {volume} {95}},\ \bibinfo
		{pages} {084014} (\bibinfo {year} {2017})},\ \Eprint
	{https://arxiv.org/abs/1701.01116} {arXiv:1701.01116 [gr-qc]} \BibitemShut
	{NoStop}%
	\bibitem [{\citenamefont {Binnington}\ and\ \citenamefont
		{Poisson}(2009)}]{BinningtonPoisson2009}%
	\BibitemOpen
	\bibfield  {author} {\bibinfo {author} {\bibfnamefont {T.}~\bibnamefont
			{Binnington}}\ and\ \bibinfo {author} {\bibfnamefont {E.}~\bibnamefont
			{Poisson}},\ }\href {https://doi.org/10.1103/PhysRevD.80.084018} {\bibfield
		{journal} {\bibinfo  {journal} {Phys. Rev. D}\ }\textbf {\bibinfo {volume}
			{80}},\ \bibinfo {pages} {084018} (\bibinfo {year} {2009})},\ \Eprint
	{https://arxiv.org/abs/0906.1366} {arXiv:0906.1366 [gr-qc]} \BibitemShut
	{NoStop}%
	\bibitem [{\citenamefont {Damour}\ and\ \citenamefont
		{Nagar}(2009)}]{DamourNagar2009}%
	\BibitemOpen
	\bibfield  {author} {\bibinfo {author} {\bibfnamefont {T.}~\bibnamefont
			{Damour}}\ and\ \bibinfo {author} {\bibfnamefont {A.}~\bibnamefont {Nagar}},\
	}\href {https://doi.org/10.1103/PhysRevD.80.084035} {\bibfield  {journal}
		{\bibinfo  {journal} {Phys. Rev. D}\ }\textbf {\bibinfo {volume} {80}},\
		\bibinfo {pages} {084035} (\bibinfo {year} {2009})},\ \Eprint
	{https://arxiv.org/abs/0906.0096} {arXiv:0906.0096 [gr-qc]} \BibitemShut
	{NoStop}%
	\bibitem [{\citenamefont {Hui}\ \emph {et~al.}(2021)\citenamefont {Hui},
		\citenamefont {Joyce}, \citenamefont {Penco}, \citenamefont {Santoni},\ and\
		\citenamefont {Solomon}}]{HuiJoycePencoSantoniSolomon2021}%
	\BibitemOpen
	\bibfield  {author} {\bibinfo {author} {\bibfnamefont {L.}~\bibnamefont
			{Hui}}, \bibinfo {author} {\bibfnamefont {A.}~\bibnamefont {Joyce}}, \bibinfo
		{author} {\bibfnamefont {R.}~\bibnamefont {Penco}}, \bibinfo {author}
		{\bibfnamefont {L.}~\bibnamefont {Santoni}},\ and\ \bibinfo {author}
		{\bibfnamefont {A.~R.}\ \bibnamefont {Solomon}},\ }\href
	{https://doi.org/10.1088/1475-7516/2021/04/052} {\bibfield  {journal}
		{\bibinfo  {journal} {JCAP}\ }\textbf {\bibinfo {volume} {2021}}\bibfield
		{number} {\bibinfo  {number} { (04)},\ \bibinfo {pages} {052}},\ }\Eprint
	{https://arxiv.org/abs/2010.00593} {arXiv:2010.00593 [hep-th]} \BibitemShut
	{NoStop}%
	\bibitem [{\citenamefont {Chia}(2021)}]{Chia2021}%
	\BibitemOpen
	\bibfield  {author} {\bibinfo {author} {\bibfnamefont {H.~S.}\ \bibnamefont
			{Chia}},\ }\href {https://doi.org/10.1103/PhysRevD.104.024013} {\bibfield
		{journal} {\bibinfo  {journal} {Phys. Rev. D}\ }\textbf {\bibinfo {volume}
			{104}},\ \bibinfo {pages} {024013} (\bibinfo {year} {2021})},\ \Eprint
	{https://arxiv.org/abs/2010.07300} {arXiv:2010.07300 [gr-qc]} \BibitemShut
	{NoStop}%
	\bibitem [{\citenamefont {{Le Tiec}}\ \emph {et~al.}(2021)\citenamefont {{Le
				Tiec}}, \citenamefont {Casals},\ and\ \citenamefont
		{Franzin}}]{LeTiecCasalsFranzin2021}%
	\BibitemOpen
	\bibfield  {author} {\bibinfo {author} {\bibfnamefont {A.}~\bibnamefont {{Le
					Tiec}}}, \bibinfo {author} {\bibfnamefont {M.}~\bibnamefont {Casals}},\ and\
		\bibinfo {author} {\bibfnamefont {E.}~\bibnamefont {Franzin}},\ }\href
	{https://doi.org/10.1103/PhysRevD.103.084021} {\bibfield  {journal} {\bibinfo
			{journal} {Phys. Rev. D}\ }\textbf {\bibinfo {volume} {103}},\ \bibinfo
		{pages} {084021} (\bibinfo {year} {2021})},\ \Eprint
	{https://arxiv.org/abs/2010.15795} {arXiv:2010.15795 [gr-qc]} \BibitemShut
	{NoStop}%
	\bibitem [{\citenamefont {Goldberger}\ and\ \citenamefont
		{Rothstein}(2006)}]{GoldbergerRothstein2006}%
	\BibitemOpen
	\bibfield  {author} {\bibinfo {author} {\bibfnamefont {W.~D.}\ \bibnamefont
			{Goldberger}}\ and\ \bibinfo {author} {\bibfnamefont {I.~Z.}\ \bibnamefont
			{Rothstein}},\ }\href {https://doi.org/10.1103/PhysRevD.73.104029} {\bibfield
		{journal} {\bibinfo  {journal} {Phys. Rev. D}\ }\textbf {\bibinfo {volume}
			{73}},\ \bibinfo {pages} {104029} (\bibinfo {year} {2006})},\ \Eprint
	{https://arxiv.org/abs/hep-th/0409156} {arXiv:hep-th/0409156 [hep-th]}
	\BibitemShut {NoStop}%
	\bibitem [{\citenamefont {Kol}\ and\ \citenamefont
		{Smolkin}(2012)}]{KolSmolkin2012}%
	\BibitemOpen
	\bibfield  {author} {\bibinfo {author} {\bibfnamefont {B.}~\bibnamefont
			{Kol}}\ and\ \bibinfo {author} {\bibfnamefont {M.}~\bibnamefont {Smolkin}},\
	}\href {https://doi.org/10.1007/JHEP02(2012)010} {\bibfield  {journal}
		{\bibinfo  {journal} {JHEP}\ }\textbf {\bibinfo {volume} {2012}}\bibfield
		{number} {\bibinfo  {number} { (02)},\ \bibinfo {pages} {010}},\ }\Eprint
	{https://arxiv.org/abs/1110.3764} {arXiv:1110.3764 [hep-th]} \BibitemShut
	{NoStop}%
	\bibitem [{\citenamefont {Hui}\ \emph {et~al.}(2022)\citenamefont {Hui},
		\citenamefont {Joyce}, \citenamefont {Penco}, \citenamefont {Santoni},\ and\
		\citenamefont {Solomon}}]{HuiJoycePencoSantoniSolomon2022}%
	\BibitemOpen
	\bibfield  {author} {\bibinfo {author} {\bibfnamefont {L.}~\bibnamefont
			{Hui}}, \bibinfo {author} {\bibfnamefont {A.}~\bibnamefont {Joyce}}, \bibinfo
		{author} {\bibfnamefont {R.}~\bibnamefont {Penco}}, \bibinfo {author}
		{\bibfnamefont {L.}~\bibnamefont {Santoni}},\ and\ \bibinfo {author}
		{\bibfnamefont {A.~R.}\ \bibnamefont {Solomon}},\ }\href
	{https://doi.org/10.1088/1475-7516/2022/01/032} {\bibfield  {journal}
		{\bibinfo  {journal} {JCAP}\ }\textbf {\bibinfo {volume} {2022}}\bibfield
		{number} {\bibinfo  {number} { (01)},\ \bibinfo {pages} {032}},\ }\Eprint
	{https://arxiv.org/abs/2105.01069} {arXiv:2105.01069 [hep-th]} \BibitemShut
	{NoStop}%
	\bibitem [{\citenamefont {Charalambous}\ \emph {et~al.}(2021)\citenamefont
		{Charalambous}, \citenamefont {Dubovsky},\ and\ \citenamefont
		{Ivanov}}]{CharalambousDubovskyIvanov2021}%
	\BibitemOpen
	\bibfield  {author} {\bibinfo {author} {\bibfnamefont {P.}~\bibnamefont
			{Charalambous}}, \bibinfo {author} {\bibfnamefont {S.}~\bibnamefont
			{Dubovsky}},\ and\ \bibinfo {author} {\bibfnamefont {M.~M.}\ \bibnamefont
			{Ivanov}},\ }\href {https://doi.org/10.1103/PhysRevLett.127.101101}
	{\bibfield  {journal} {\bibinfo  {journal} {Phys. Rev. Lett.}\ }\textbf
		{\bibinfo {volume} {127}},\ \bibinfo {pages} {101101} (\bibinfo {year}
		{2021})},\ \Eprint {https://arxiv.org/abs/2103.01234} {arXiv:2103.01234
		[hep-th]} \BibitemShut {NoStop}%
	\bibitem [{\citenamefont {Ben~Achour}\ \emph {et~al.}(2022)\citenamefont
		{Ben~Achour}, \citenamefont {Livine}, \citenamefont {Mukohyama},\ and\
		\citenamefont {Uzan}}]{BenAchourLivineMukohyamaUzan2022}%
	\BibitemOpen
	\bibfield  {author} {\bibinfo {author} {\bibfnamefont {J.}~\bibnamefont
			{Ben~Achour}}, \bibinfo {author} {\bibfnamefont {E.~R.}\ \bibnamefont
			{Livine}}, \bibinfo {author} {\bibfnamefont {S.}~\bibnamefont {Mukohyama}},\
		and\ \bibinfo {author} {\bibfnamefont {J.-P.}\ \bibnamefont {Uzan}},\ }\href
	{https://doi.org/10.1007/JHEP07(2022)112} {\bibfield  {journal} {\bibinfo
			{journal} {JHEP}\ }\textbf {\bibinfo {volume} {2022}}\bibfield  {number}
		{\bibinfo  {number} { (07)},\ \bibinfo {pages} {112}},\ }\Eprint
	{https://arxiv.org/abs/2202.12828} {arXiv:2202.12828 [gr-qc]} \BibitemShut
	{NoStop}%
	\bibitem [{\citenamefont {Riva}\ \emph {et~al.}(2024)\citenamefont {Riva},
		\citenamefont {Santoni}, \citenamefont {Savi{\'c}},\ and\ \citenamefont
		{Vernizzi}}]{RivaSantoniSavicVernizzi2024}%
	\BibitemOpen
	\bibfield  {author} {\bibinfo {author} {\bibfnamefont {M.~M.}\ \bibnamefont
			{Riva}}, \bibinfo {author} {\bibfnamefont {L.}~\bibnamefont {Santoni}},
		\bibinfo {author} {\bibfnamefont {N.}~\bibnamefont {Savi{\'c}}},\ and\
		\bibinfo {author} {\bibfnamefont {F.}~\bibnamefont {Vernizzi}},\ }\href
	{https://doi.org/10.1016/j.physletb.2024.138710} {\bibfield  {journal}
		{\bibinfo  {journal} {Phys. Lett. B}\ }\textbf {\bibinfo {volume} {854}},\
		\bibinfo {pages} {138710} (\bibinfo {year} {2024})},\ \Eprint
	{https://arxiv.org/abs/2312.05065} {arXiv:2312.05065 [gr-qc]} \BibitemShut
	{NoStop}%
	\bibitem [{\citenamefont {Endlich}\ \emph {et~al.}(2017)\citenamefont
		{Endlich}, \citenamefont {Gorbenko}, \citenamefont {Huang},\ and\
		\citenamefont {Senatore}}]{EndlichGorbenkoHuangSenatore2017}%
	\BibitemOpen
	\bibfield  {author} {\bibinfo {author} {\bibfnamefont {S.}~\bibnamefont
			{Endlich}}, \bibinfo {author} {\bibfnamefont {V.}~\bibnamefont {Gorbenko}},
		\bibinfo {author} {\bibfnamefont {J.}~\bibnamefont {Huang}},\ and\ \bibinfo
		{author} {\bibfnamefont {L.}~\bibnamefont {Senatore}},\ }\href
	{https://doi.org/10.1007/JHEP09(2017)122} {\bibfield  {journal} {\bibinfo
			{journal} {JHEP}\ }\textbf {\bibinfo {volume} {2017}}\bibfield  {number}
		{\bibinfo  {number} { (09)},\ \bibinfo {pages} {122}},\ }\Eprint
	{https://arxiv.org/abs/1704.01590} {arXiv:1704.01590 [gr-qc]} \BibitemShut
	{NoStop}%
	\bibitem [{\citenamefont {Cardoso}\ \emph {et~al.}(2018)\citenamefont
		{Cardoso}, \citenamefont {Kimura}, \citenamefont {Maselli},\ and\
		\citenamefont {Senatore}}]{CardosoKimuraMaselliSenatore2018}%
	\BibitemOpen
	\bibfield  {author} {\bibinfo {author} {\bibfnamefont {V.}~\bibnamefont
			{Cardoso}}, \bibinfo {author} {\bibfnamefont {M.}~\bibnamefont {Kimura}},
		\bibinfo {author} {\bibfnamefont {A.}~\bibnamefont {Maselli}},\ and\ \bibinfo
		{author} {\bibfnamefont {L.}~\bibnamefont {Senatore}},\ }\href
	{https://doi.org/10.1103/PhysRevLett.121.251105} {\bibfield  {journal}
		{\bibinfo  {journal} {Phys. Rev. Lett.}\ }\textbf {\bibinfo {volume} {121}},\
		\bibinfo {pages} {251105} (\bibinfo {year} {2018})},\ \bibinfo {note}
	{erratum: Phys. Rev. Lett. \textbf{131}, 109903 (2023),
		doi:10.1103/PhysRevLett.131.109903},\ \Eprint
	{https://arxiv.org/abs/1808.08962} {arXiv:1808.08962 [gr-qc]} \BibitemShut
	{NoStop}%
	\bibitem [{\citenamefont {Sennett}\ \emph {et~al.}(2020)\citenamefont
		{Sennett}, \citenamefont {Brito}, \citenamefont {Buonanno}, \citenamefont
		{Gorbenko},\ and\ \citenamefont
		{Senatore}}]{SennettBritoBuonannoGorbenkoSenatore2020}%
	\BibitemOpen
	\bibfield  {author} {\bibinfo {author} {\bibfnamefont {N.}~\bibnamefont
			{Sennett}}, \bibinfo {author} {\bibfnamefont {R.}~\bibnamefont {Brito}},
		\bibinfo {author} {\bibfnamefont {A.}~\bibnamefont {Buonanno}}, \bibinfo
		{author} {\bibfnamefont {V.}~\bibnamefont {Gorbenko}},\ and\ \bibinfo
		{author} {\bibfnamefont {L.}~\bibnamefont {Senatore}},\ }\href
	{https://doi.org/10.1103/PhysRevD.102.044056} {\bibfield  {journal} {\bibinfo
			{journal} {Phys. Rev. D}\ }\textbf {\bibinfo {volume} {102}},\ \bibinfo
		{pages} {044056} (\bibinfo {year} {2020})},\ \Eprint
	{https://arxiv.org/abs/1912.09917} {arXiv:1912.09917 [gr-qc]} \BibitemShut
	{NoStop}%
	\bibitem [{\citenamefont {Cano}\ \emph {et~al.}(2022)\citenamefont {Cano},
		\citenamefont {Ganchev}, \citenamefont {Mayerson},\ and\ \citenamefont
		{Ruip{\'e}rez}}]{CanoGanchevMayersonRuiperez2022}%
	\BibitemOpen
	\bibfield  {author} {\bibinfo {author} {\bibfnamefont {P.~A.}\ \bibnamefont
			{Cano}}, \bibinfo {author} {\bibfnamefont {B.}~\bibnamefont {Ganchev}},
		\bibinfo {author} {\bibfnamefont {D.~R.}\ \bibnamefont {Mayerson}},\ and\
		\bibinfo {author} {\bibfnamefont {A.}~\bibnamefont {Ruip{\'e}rez}},\ }\href
	{https://doi.org/10.1007/JHEP12(2022)120} {\bibfield  {journal} {\bibinfo
			{journal} {JHEP}\ }\textbf {\bibinfo {volume} {2022}}\bibfield  {number}
		{\bibinfo  {number} { (12)},\ \bibinfo {pages} {120}},\ }\Eprint
	{https://arxiv.org/abs/2208.01044} {arXiv:2208.01044 [gr-qc]} \BibitemShut
	{NoStop}%
	\bibitem [{\citenamefont {Teukolsky}(1973)}]{Teukolsky1973}%
	\BibitemOpen
	\bibfield  {author} {\bibinfo {author} {\bibfnamefont {S.~A.}\ \bibnamefont
			{Teukolsky}},\ }\href {https://doi.org/10.1086/152444} {\bibfield  {journal}
		{\bibinfo  {journal} {Astrophys. J.}\ }\textbf {\bibinfo {volume} {185}},\
		\bibinfo {pages} {635} (\bibinfo {year} {1973})}\BibitemShut {NoStop}%
	\bibitem [{\citenamefont {Cano}(2025)}]{Cano2025}%
	\BibitemOpen
	\bibfield  {author} {\bibinfo {author} {\bibfnamefont {P.~A.}\ \bibnamefont
			{Cano}},\ }\href {https://doi.org/10.1007/JHEP07(2025)152} {\bibfield
		{journal} {\bibinfo  {journal} {JHEP}\ }\textbf {\bibinfo {volume}
			{2025}}\bibfield  {number} {\bibinfo  {number} { (07)},\ \bibinfo {pages}
			{152}},\ }\Eprint {https://arxiv.org/abs/2502.20185} {arXiv:2502.20185
		[gr-qc]} \BibitemShut {NoStop}%
	\bibitem [{\citenamefont {Ali~Barura}\ \emph {et~al.}(2024)\citenamefont
		{Ali~Barura}, \citenamefont {Kobayashi}, \citenamefont {Mukohyama},
		\citenamefont {Oshita}, \citenamefont {Takahashi},\ and\ \citenamefont
		{Yingcharoenrat}}]{GharibBaruraEtAl2024}%
	\BibitemOpen
	\bibfield  {author} {\bibinfo {author} {\bibfnamefont {C.~G.}\ \bibnamefont
			{Ali~Barura}}, \bibinfo {author} {\bibfnamefont {H.}~\bibnamefont
			{Kobayashi}}, \bibinfo {author} {\bibfnamefont {S.}~\bibnamefont
			{Mukohyama}}, \bibinfo {author} {\bibfnamefont {N.}~\bibnamefont {Oshita}},
		\bibinfo {author} {\bibfnamefont {K.}~\bibnamefont {Takahashi}},\ and\
		\bibinfo {author} {\bibfnamefont {V.}~\bibnamefont {Yingcharoenrat}},\
	}\href@noop {} {\  (\bibinfo {year} {2024})},\ \Eprint
	{https://arxiv.org/abs/2405.10813} {arXiv:2405.10813 [gr-qc]} \BibitemShut
	{NoStop}%
	\bibitem [{\citenamefont {Barbosa}\ \emph {et~al.}(2025)\citenamefont
		{Barbosa}, \citenamefont {Brax}, \citenamefont {Fichet},\ and\ \citenamefont
		{de~Souza}}]{BarbosaBraxFichetDeSouza2025}%
	\BibitemOpen
	\bibfield  {author} {\bibinfo {author} {\bibfnamefont {S.}~\bibnamefont
			{Barbosa}}, \bibinfo {author} {\bibfnamefont {P.}~\bibnamefont {Brax}},
		\bibinfo {author} {\bibfnamefont {S.}~\bibnamefont {Fichet}},\ and\ \bibinfo
		{author} {\bibfnamefont {L.}~\bibnamefont {de~Souza}},\ }\href@noop {} {\
		(\bibinfo {year} {2025})},\ \Eprint {https://arxiv.org/abs/2501.18684}
	{arXiv:2501.18684 [gr-qc]} \BibitemShut {NoStop}%
	\bibitem [{\citenamefont {Katagiri}\ \emph {et~al.}(2024)\citenamefont
		{Katagiri}, \citenamefont {Ikeda},\ and\ \citenamefont
		{Cardoso}}]{KatagiriIkedaCardoso2024}%
	\BibitemOpen
	\bibfield  {author} {\bibinfo {author} {\bibfnamefont {T.}~\bibnamefont
			{Katagiri}}, \bibinfo {author} {\bibfnamefont {T.}~\bibnamefont {Ikeda}},\
		and\ \bibinfo {author} {\bibfnamefont {V.}~\bibnamefont {Cardoso}},\ }\href
	{https://doi.org/10.1103/PhysRevD.109.044067} {\bibfield  {journal} {\bibinfo
			{journal} {Phys. Rev. D}\ }\textbf {\bibinfo {volume} {109}},\ \bibinfo
		{pages} {044067} (\bibinfo {year} {2024})},\ \Eprint
	{https://arxiv.org/abs/2311.00237} {arXiv:2311.00237 [gr-qc]} \BibitemShut
	{NoStop}%
	\bibitem [{\citenamefont {Wang}\ \emph {et~al.}(2026)\citenamefont {Wang},
		\citenamefont {Lehner}, \citenamefont {Micol},\ and\ \citenamefont
		{Sturani}}]{WangLehnerMicolSturani2026}%
	\BibitemOpen
	\bibfield  {author} {\bibinfo {author} {\bibfnamefont {L.}~\bibnamefont
			{Wang}}, \bibinfo {author} {\bibfnamefont {L.}~\bibnamefont {Lehner}},
		\bibinfo {author} {\bibfnamefont {M.}~\bibnamefont {Micol}},\ and\ \bibinfo
		{author} {\bibfnamefont {R.}~\bibnamefont {Sturani}},\ }\href@noop {} {\
		(\bibinfo {year} {2026})},\ \Eprint {https://arxiv.org/abs/2604.04259}
	{arXiv:2604.04259 [gr-qc]} \BibitemShut {NoStop}%
	\bibitem [{\citenamefont {Regge}\ and\ \citenamefont
		{Wheeler}(1957)}]{ReggeWheeler1957}%
	\BibitemOpen
	\bibfield  {author} {\bibinfo {author} {\bibfnamefont {T.}~\bibnamefont
			{Regge}}\ and\ \bibinfo {author} {\bibfnamefont {J.~A.}\ \bibnamefont
			{Wheeler}},\ }\href {https://doi.org/10.1103/PhysRev.108.1063} {\bibfield
		{journal} {\bibinfo  {journal} {Phys. Rev.}\ }\textbf {\bibinfo {volume}
			{108}},\ \bibinfo {pages} {1063} (\bibinfo {year} {1957})}\BibitemShut
	{NoStop}%
	\bibitem [{\citenamefont {Zerilli}(1970)}]{Zerilli1970}%
	\BibitemOpen
	\bibfield  {author} {\bibinfo {author} {\bibfnamefont {F.~J.}\ \bibnamefont
			{Zerilli}},\ }\href {https://doi.org/10.1103/PhysRevLett.24.737} {\bibfield
		{journal} {\bibinfo  {journal} {Phys. Rev. Lett.}\ }\textbf {\bibinfo
			{volume} {24}},\ \bibinfo {pages} {737} (\bibinfo {year} {1970})}\BibitemShut
	{NoStop}%
\end{thebibliography}
\end{document}